\documentclass[11pt]{article}
\usepackage{amsbsy}
\newcommand{\be}{\begin{equation}}
\newcommand{\ee}{\end{equation}}
\newcommand{\bos}{\boldsymbol}
\newcommand{\R}{\mbox{\scriptsize I\hspace{-.19em}R}}
\newcommand{\Rbig}{\mbox{I\hspace{-.19em}R}}

\newtheorem{result}{Result}[section] 
\newtheorem{remark}{Remark}[section]

\newcommand{\ab}{\mbox{\bf A}}
\newcommand{\bb}{\mbox{\bf B}}
\newcommand{\ub}{\mbox{\bf U}}
\newcommand{\eb}{\mbox{\bf E}}
\newcommand{\ob}{\mbox{\bf O}}
\newcommand{\f}{\mbox{\bf f}}
\newcommand{\A}{{\cal{A}}}
\newcommand{\B}{{\cal{B}}}
\newcommand{\C}{{\cal{C}}}
\newcommand{\Pc}{{\cal{P}}}
\renewcommand{\theequation}{\thesection.\arabic{equation}}
\begin{document}
\title{\Large\bf 
 Symbolic calculus on the time-frequency half-plane}
\par
\vspace{5mm}
\author{{\it   J.Bertrand$^{\ast }$ and P.Bertrand$^{\ast \ast}$}     
\\ [3mm]
\normalsize ($^{\ast }$) LPTM, University Paris VII, 
75251 Paris Cedex 05, France \\
\normalsize ($^{\ast \ast}$) ONERA, Chemin de la Huni\`ere, F-91761,
Palaiseau Cedex, France }
\vspace{3mm}
\date{}
\maketitle
\par
\vspace{2cm}
 \begin{center}
{\large\bf Abstract}
\end{center}
The study concerns a special symbolic calculus of interest for signal analysis. 
This calculus associates functions on the time-frequency half-plane $f>0$ with linear operators 
defined on the positive-frequency signals. Full attention is given to its construction 
which is entirely based on the study of the affine group in a simple and direct way. 
 The correspondence rule is 
detailed and the associated Wigner function is given.  
Formulas  expressing the basic operation (star-bracket) of the Lie algebra of symbols, 
which is isomorphic to that of the operators, are obtained. 
In addition, it is shown that the resulting calculus is covariant under a three-parameter 
group which contains the affine group as subgroup. This observation is the starting point 
of an investigation leading to a whole class of symbolic calculi which can be considered 
as modifications of the original one. 
 \par
 
\vspace{1.6cm}

\section{Introduction}

The notion of time-frequency analysis of signals received a theoretical basis 
when  D.Gabor 
\cite{gabor} proposed to interpret signals by operations very similar to those 
of quantum mechanics.
In this approach, the real signals are replaced by their positive-frequency parts (the 
"analytic" signals) and the communication characteristics are introduced as the 
mean values of hermitian operators defined on the corresponding Hilbert space. 
Actually, the analogy with quantum mechanics has rapidly suggested to recognize the time and 
the frequency as two conjugate variables and has led to look for a phase 
space formulation of the theory. Various solutions have then been proposed which were essentially 
adaptations of solutions developed precedently in the quantum mechanical context.
\par
From a mathematical point of view, the question of the phase space formulation of a quantum 
theory appears as a problem of symbolic calculus since each operator of the theory has 
to be replaced by a function (its symbol) of the phase space variables 
 \cite{flato}.
The construction of such a correspondence rule is determined by the class of operators 
to symbolize and by subsidiary constraints relative to the physical interpretation of 
the formalism. In signal theory, this last point concerns simply the formal invariance of 
the operations in changes of reference clocks. 
The corresponding constraint will be automatically ensured if the correspondence rule is covariant with
respect to relevant representations of the affine group. The object of the following developments is to construct 
a symbolic calculus on the time-frequency half-plane along these lines.
\par

Section 2 gives a study of the affine group and its subgroups, their representations in the 
Hilbert space ${\cal H}$ of signals and their actions in the time-frequency 
half-plane $\Gamma$. Decompositions of ${\cal H}$ and $\Gamma$ into invariant 
subspaces are carried out in parallel and shown to be related in a natural way. 
The correspondence rule is derived in 
Section 3 by exploiting the completeness of the invariant decompositions just performed. 
Some properties of this rule are given in Section 4 and the associated Wigner function is determined. Section 5 
is devoted to the derivation of the star product, which allows to symbolize products of operators and 
is the core of the symbolic calculus. In Section 6, the star bracket is introduced and shown to reduce 
to an ordinary Poisson bracket in some special cases. As an application of this remark, a Hamiltonian flow is 
defined in phase space. Finally, it is shown in Section 7 that the method has possible 
extensions when working with 
larger groups containing the affine group.
\par

\section{The affine group in signal theory}{\label{sec1}}
\setcounter{equation}{0}

The signals to be considered are real-valued functions of time $s(t)$
which are interpreted independently of the choice of time 
origin and units. This means that the description of the signal processing 
operations must not change if time $t$ is replaced by 
$at+b$ 
where $b$ is real and $a$ positive. 
The corresponding change 
 on the signal can be readily written down. Notice that there is the freedom of 
rescaling the signal when the units are changed. The 
transformations on the signal that will be considered are written as:
\be
s(t)\longrightarrow a^r \,s(a^{-1}(t-b))
 \label{reals}
\ee
\noindent where $r$ is a real exponent.\par
 
The set of transformations (\ref{reals})  forms the affine group, 
with parameters $(a,b)$ and composition law expressed by:
\be
(a,b)(a',b') =(aa', b+ab')
\ee
The group structure is the mathematical expression of the equivalence of the 
reference clocks.
To stress the fact that only positive dilations are considered, 
the parameter $u$ such that:
\be
a=e^u 
\label{u}
\ee
 will  be used instead of $a$.
\vfill\eject

\subsection{Representation in the Hilbert space of positive-frequency signals}\label{21}

The Fourier transform of the 
signal $s(t)$ will be introduced by:
\be
\hat{s}(f) \equiv \int_{\R} s(t) e^{-2i\pi ft} \; dt
\label{fourier}
\ee
Since $s(t)$ is a real-valued function, 
it is   characterized  by 
its positive frequency part only. 
In agreement with the usual practice in signal theory, this property 
will be used to substitute to $s(t)$ the so-called analytic signal 
whose Fourier transform is given by:   
\be
S(f)= 2 \, Y(f) \hat{s}(f)
\ee
\noindent where $Y(f) $ is Heaviside's step function.\par 
The counterpart of transformation (\ref{reals}) for the analytic signal $S(f)$
is easily obtained. 
Namely, by action of an element $(u,b)$ of the 
affine group, signal $S(f)$ goes to a new function 
$ [\ub (u,b)S](f)$ defined by:
 
\be
 \ub (u,b)S(f) \equiv e^{(r+1)u}\, e^{-2i\pi fb}\, S(e^uf), \quad r \in \Rbig
\label{uab}
\ee

The operators $\ub (u,b)$ constitute an irreducible representation of the affine group  
that is unitary for the scalar product defined by
\be
(S,S') \equiv \int_{\R^+} S(f) S'^*(f) \, f^{2r+1} \, df
\label{scalar}
\ee

The Hilbert space of functions $S(f)$ on the half-line 
which are square integrable for the measure $f^{2r+1} \, df  $
will   be denoted by ${\cal{H}}$ (the reference to $r$ being 
implicit).  
From a physical point of view, it may be necessary to 
distinguish between signals transforming according to 
(\ref{uab}) with different values of $r$. However, the 
corresponding representations are easily shown to be unitarily 
equivalent.
In fact, it may be shown that the affine group has only two inequivalent 
unitary representations differing by the sign of the imaginary 
exponential in (\ref{uab}) \cite{vilenkin}.

The infinitesimal operators of representation (\ref{uab}) are  introduced  by 
the relations:
 \begin{eqnarray}
{\boldsymbol{\beta}}S(f) & \equiv & - 
\frac{1}{2i\pi} \left. \frac{d}{du}\ub(u,b)S(f) \right|_{u=b=0}  
\nonumber \\[2mm]
& = & -\frac{1}{2i\pi}\left( r+1+f\frac{d}{df} \right) S(f) 
 \label{genb} \\[2mm]
\f S(f) & \equiv & -
\frac{1}{2i\pi} \left. \frac{d}{db}\ub(u,b)S(f) \right|_{u=b=0}
\nonumber \\[2mm]
& = & fS(f)
\label{genf}
\end{eqnarray}
The operators  $\boldsymbol{\beta}$ and $ \f$ thus obtained verify the commutation 
relation:
\be
[\boldsymbol{  \beta}, \f \, ] = - \frac{1}{2i\pi} \f
\ee
The subgroups of the affine group and their representations will now be studied in detail.
There are two kinds of   subgroups  
   consisting either of  pure dilations   or of pure translations. 
The translation subgroup is invariant by conjugation. On the contrary, since 
a dilation can be performed from an arbitrary time origin $\xi$, there is a whole family 
of conjugate dilation subgroups $G_{\xi}$ which 
consist of the following set of elements:
  
\be
G_{\xi} = \{(e^u, \xi (1-e^u))\}
\label{gxi}
\ee

The restriction ${\ub }_{\xi}$ of representation  (\ref{uab}) to $G_{\xi}$
is given by
\be
{\ub }_{\xi}(u)S(f) = e^{(r+1)u} \, e^{-2i\pi f \xi (1-e^u) } S(e^uf)
\label{uxi}
\ee
   
and the corresponding generator is
\be
\bos{\beta}_{\xi} S(f) \equiv - 
\frac{1}{2i\pi} \left. \frac{d}{du}\ub_{\xi}(u)S(f) \right|_{u=0} 
=( \bos{\beta} - \xi \bos{f})S(f)
\label{betaxi}
\ee
where operators $\bos{\beta}$ and $\bos{f}$ are given by  
(\ref{genb}) and (\ref{genf}) 
respectively.\par
The operator of representation (\ref{uxi}) can be written in terms of generator
 ${\bos \beta }_{\xi}$ as:
\be
\ub _{\xi} = e^{-2i\pi u { \bos {\beta}}_{\xi}}
\label{ubetaxi}
\ee

For a given $\xi$, the eigenfunctions of the self-adjoint operator 
$\bos{\beta}_{\xi}$   
  are defined by the equation
\be
\bos{\beta}_{\xi} \psi^{\xi}_{\beta}(f) =\beta \psi^{\xi}_{\beta}(f) , \quad \beta \in \Rbig
\ee
\noindent whose solution is, up to a multiplicative constant:  
 \be
\psi^{\xi}_{\beta}(f) 
\equiv f^{-2i\pi \beta -r-1} \, e^{-2i\pi \xi f}
\label{psi}
\ee
By construction, the functions $\psi^{\xi}_{\beta}(f)$ are also 
eigenfunctions of the unitary operator (\ref{ubetaxi}) and verify
\be
{\ub }_{\xi}(u) \psi^{\xi}_{\beta}(f) = e^{-2i\pi \beta u } \, \psi^{\xi}_{\beta}(f) 
\label{prop4}
\ee
For a given $\xi$, functions (\ref{psi}) constitute an improper basis of the Hilbert space  ${\cal H}$.
In fact, one verifies that there is 
 orthogonality of the functions corresponding to 
 different $\beta$ 
 \be
(\psi^{\xi}_{\beta}(f), 
\psi^{\xi}_{\beta'}(f)) = \delta (\beta -\beta') 
\label{orth}
\ee
\noindent 
where the scalar product has the form (\ref{scalar}).
The   completeness of this basis is expressed by:
\be
\int_{\R} 
\psi^{\xi}_{\beta}(f) 
\psi^{\xi *}_{\beta}(f')  \; d\beta =
f^{-2r-1} \, \delta (f-f')
\ee
For each $\xi$, the set of all the projectors on the functions 
$  \psi^{\xi}_{\beta} (f) $ gives a decomposition of the Hilbert space ${\cal H}$. 
From (\ref{prop4}) it results that each of these projectors 
is invariant by the transformations of the subgroup $G_{\xi}$. \par
More generally, the action of the whole affine group on functions 
$  \psi^{\xi}_{\beta} (f) $ is given by:
\be
\ub (u,b) \psi^{\xi}_{\beta}(f) = e^{-2i\pi u \beta} \, \psi^{\xi'}_{\beta}(f)
\label{usurpsi}
\ee
\noindent with $\xi' = e^u \xi + b$.
This shows that the set of projectors associated with subgroup $G_{\xi}$ 
is transformed as a whole into the set of projectors associated 
with subgroup  $G_{e^u\xi +b}$. In other words, this means that the set of all 
decompositions of ${\cal H}$ by the various subgroups  $G_{\xi}$  is invariant 
by action of the whole group. 
\subsection{Action of the group in the time-frequency half-plane }

The physical phase space for signal analysis is the 
time-frequency half-plane, $\Gamma = \{ (t,f) \}$, 
 $ t$ real, $f>0$.
This space can also be obtained as an orbit of the coadjoint representation 
of the affine group whose action in variables $(t,f)$ is given by:
 \be
(u,b): \quad (t,f) \longrightarrow (e^ut+b,e^{-u} f)
\label{coad}
\ee
\noindent where the pair $(u,b)$ characterizes an element of the group. \par
 This action induces a transformation on  functions $\A(t,f)$ defined on $\Gamma$. 
This transformation has the form: 
\be
(u,b): \A(t,f) \longrightarrow \A(e^{-u}(t-b),e^uf)
\ee
According to definition (\ref{gxi}), the restriction of  
 representation (\ref{coad}) to subgroup $G_{\xi}$ reads:
\be
 (t,f) \longrightarrow (e^ut+ \xi(1-e^u), e^{-u}f)
\label{actf}
\ee
and the corresponding transformation of phase space functions is given by:
\be
 \A(t,f) \longrightarrow \A(e^{-u}(t-\xi)+ \xi, e^uf)
\label{aeu}
\ee
The transformation (\ref{actf}) has the following invariant:
\be
(t-\xi )f = \tilde{\beta},\quad \tilde{\beta} \in \Rbig
\label{curve}
\ee
\noindent where $\tilde{\beta}$ is an arbitrary  constant. For each $\xi$, this 
relation defines a family of curves in 
phase space  labelled by $\tilde{\beta}$. Actually    any point 
in $\Gamma$ belongs to one, and only one, of these curves and a partition of 
phase space has thus been achieved for any value of $\xi$. 
\par
Uniform distributions on curves (\ref{curve}) are given by the Dirac 
distributions:
\be
\delta_{\beta}^{\xi} \equiv \delta ((t-\xi )f -\tilde{\beta} )
\ee
These distributions are invariant by action of the subgroup $G_{\xi}$ and transform 
under the 
full group as follows:
\be
(u,b): \; \delta_{\tilde{\beta}}^{\xi} = \delta ((t-\xi )f -\tilde{\beta} ) \longrightarrow 
\delta ((e^{-u}(t-b)-\xi )e^uf -\tilde{\beta} )=\delta_{\tilde{\beta}}^{\xi e^u+b} 
\ee
This result can be compared with (\ref{usurpsi}). In fact, the full group connects the 
various phase space partitions associated with subgroups $G_{\xi}$. 

\subsection{Correspondence between invariant structures}\label{23}

The subgroup $G_{\xi}$, considered for any given value of $\xi$, has 
allowed to perform decompositions of both $\cal{H}$ and $\Gamma$
where the resulting partitions have been labelled by parameters denoted 
 respectively by $\beta$ and $\tilde{\beta}$. We will now show that 
a natural connection does exist between these two parameters. 
To develop this point, it is useful to introduce  the notion 
of localized signals in a consistent way.
\par
 A signal $S_{t_0}(f)\in {\cal H}$ is said to be localized at time $t_0$ 
provided an affine transformation, defined by the element 
$(u,b)$ of the affine group, sends it into a signal localized at $(e^ut_0 +b)$. 
This is expressed by the following equation: 
 \be
{\ub }(u,b)S_{t_0}(f) \equiv 
e^{(r+1)u} \, e^{-2i\pi fb} \, S_{t_0}(e^uf)= 
S_{e^ut_0+b}(f)
\ee
whose solution is, up to a multiplicative constant:
\be
S_{t_0}(f) = f^{-r-1} \, e^{-2i\pi f t_0}
\label{loc}
\ee
Notice the factor $f^{-r-1}$ which is required by dilation invariance. 
In this way,  the whole set of localized states $S_{t_0}$ is stable by action of the affine group. 
\par

In the decomposition of ${\cal H}$ relative to the subgroup $G_{\xi}$, the $\beta$-subspace 
is  
generated by 
  $\psi^{\xi}_{\beta}(f)$ of the form (\ref{psi}). Developing the phase of this function 
in the vicinity of a frequency $f_0$ leads to the approximate expression:
\be
\psi^{\xi}_{\beta}(f)
 \simeq e^{-2i\pi \beta (\ln f_0 -1)} \, f^{-r-1} \, e^{-2i\pi f(\beta /f_0 +\xi)} 
\label{micro}
\ee
A comparison with (\ref{loc}) shows that, apart from a constant phase, the 
approximated function  can be seen as a state localized at:
\be
t_0 = \beta /f_0 + \xi 
\label{bb}
\ee
Thus, in a local study of functions (\ref{psi}),  it is possible to associate with any frequency $f$ a time $t$ 
given by $t=\beta /f + \xi$. 
This relation defines precisely an orbit of subgroup 
$G_{\xi}$ in phase space $\Gamma$ (cf.(\ref{curve})). It leads naturally  to identify the 
 parameters of the connected elements  by setting:
\be
  \tilde{\beta}=\beta
\ee
This identification, which is clearly invariant under the group action, 
will play a central role in the following construction.

\section{Geometric correspondence between symbols and operators}
\renewcommand{\theequation}{\thesection.\arabic{equation}}
\setcounter{equation}{0}

The aim  now is to set up a correspondence between 
operators on the Hilbert space $\cal{H}$ and functions (their symbols) on the 
time-frequency half-plane $\Gamma $. 
The approach will be essentially based on the parallel decompositions of 
$\cal{H}$ and $\Gamma$ introduced in the preceding section.   
 The derived correspondence will be said to be geometric to  emphasize the fact 
that it is essentially founded on a study of the group and of the invariants 
associated with its subgroups. \par
 
 \par
\subsection{Characterization of operators by their diagonal elements in the various
 ${\bos \xi}$-bases}
Consider an   operator $\ab$ defined by a kernel according to:
\be 
[{\bos A }S](f_1) = \int_{\R^+} A (f_1,f_2) S(f_2) f_2^{2r+1} \; df_2
\label{kernel}
\ee
	
Its diagonal matrix elements on the $G_{\xi}$-invariant basis 
defined by (\ref{psi}) are given by:
\begin{eqnarray}
I_{\cal{H}}(\xi ,\beta ) & \equiv & 
(\psi_{\beta}^{\xi}, \ab \psi_{\beta}^{\xi}) \label{ih} \\[2mm]
 & & = \int_0^{\infty} \int_0^{\infty} 
\psi_{\beta}^{\xi *}(f_1) A(f_1,f_2) \psi_{\beta}^{\xi}(f_2)
(f_1f_2)^{2r+1} \; df_1 df_2
\nonumber
\end{eqnarray}
 
Quantities $I_{\cal{H}}(\xi ,\beta )$ have  several properties that will be detailed 
below. First, for any given value of 
$\xi $ and $\beta$, they are invariant by action of the representation 
${\bos U}_{\xi}$ on the operator $\ab$. Moreover, the 
operator $\ab$ can be recovered from the values of  
$I_{\cal{H}}(\xi ,\beta )$ ($-\infty< \xi <\infty ,  \, -\infty< \beta <\infty $)  
   and  a simple formula exists for the trace of operators.
To be able to discriminate between functions $I_{\cal{H}}(\xi ,\beta )$ 
associated with distinct operators, we will also use the notation  
$I_{\cal{H}}(\ab ;\xi ,\beta )$ for (\ref{ih}).
\begin{description}
\item
{\em $G_{\xi}$-invariance of $I_{\cal{H}}(\xi ,\beta )$  }\par
The matrix elements of the transformed operator $\ab'={\ub}_{\xi}^{-1} \, \ab\,\ub_{\xi}$ 
are given by : 
\begin{eqnarray*}
(\psi_{\beta}^{\xi},\ub_{\xi}^{-1} \, \ab \,\ub_{\xi} \psi_{\beta}^{\xi}) & = & 
({ \ub}_{\xi}\psi_{\beta}^{\xi}, \ab \ub_{\xi} \psi_{\beta}^{\xi}) \\[2mm]
& = & 
(\psi_{\beta}^{\xi}, \ab \psi_{\beta}^{\xi})
\end{eqnarray*}
the last equality being a consequence of property (\ref{prop4}).
\item
{\em Affine covariance}\par
The action of the full group on $I_{\cal{H}}(\xi ,\beta )$ 
is computed in the same way, using 
property (\ref{usurpsi}). One obtains:
\be
I_{\cal{H}}( \ab';\xi ,\beta ) = I_{\cal{H}}(\ab ;e^u \xi +b ,\beta )
\label{invih}
\ee
\noindent where $\ab' =  \ub^{-1}(u,b) \ab \ub(u,b)$.\par
This relation expresses the diagonal elements of $\ab'$ in the $\xi$-basis in terms of the diagonal 
elements of $\ab$ in the basis corresponding to the subgroup $G_{\xi'}$, conjugate of 
$G_{\xi}$ by the transformation $(u,b)$ of the affine group. 
\item
{\em Reconstruction of the operator $\ab$} \par
Defining the two-dimensional Fourier transform of $I_{\cal{H}}( \xi , \beta )$ by
\be
\widehat{I}_{\cal{H}}(u,v) \equiv \int_{\R^2} e^{-2i\pi (u\xi +v\beta )} 
I_{\cal{H}}(\xi ,\beta ) \; d\xi d\beta
\label{ff}
\ee
\noindent we can compute it from (\ref{ih}) and find
\be
\widehat{I}_{\cal{H}}(u,v) = 
A\left(\frac{ue^{(v/2)}}{2\sinh (v/2)}, 
\frac{ue^{(-v/2)}}{2\sinh (v/2)} \right) 
\left( \frac{u}{2\sinh (v/2)} \right) ^{2r} 
\frac{|u|}{4\sinh^2 (v/2)} \, Y(u/v)
\label{hatih}
\ee
The inversion of this relation can be done directly by changing the variables. 
The kernel of operator $\ab$ is thus given by:
\be
A (f_1,f_2) = \widehat{I}_{\cal{H}}(f_1-f_2, \ln (f_1/f_2)) (f_1f_2)^{-r-1}
|f_1-f_2|
\label{invh}
\ee
\item  
{\em Trace of a product of two operators} \par
The scalar product of two operators can be defined as
\be
\mbox{Tr} \,( \ab{\bb}^{\dagger}) = 
\int_{\R^+ \times \R^+} A (f_1,f_2) B^*(f_1,f_2) (f_1f_2)^{2r+1} \; df_1df_2
\label{trace}
\ee
\noindent where $\bb^{\dagger}$ is the hermitian conjugate of operator $\bb$. \par
If $\widehat{I}_{\cal{H}}(\ab; u,v)$
and $\widehat{I}_{\cal{H}}({\bb}; u,v)$  are the functions 
corresponding respectively to the operators $\ab$ and ${\bb}$, 
the computation of the trace of $\ab{\bb}^{\dagger}$ is performed 
 using (\ref{invh}) and the result follows:
 \be
\mbox{Tr} \, (\ab{\bb}^{\dagger})
=\int_{\R^2}  \widehat{I}_{\cal{H}}(\ab; u,v) 
\widehat{I}_{\cal{H}}({\bb^*}; u,v) |u| \; dudv
\label{u1}
\ee

\end{description}

\subsection{Characterization of symbols by their line integrals on ${\bos \xi }$-orbits}

The description of functions ${\cal{A}}(t,f)$ defined on  phase space $\Gamma $ can 
be carried out in a way which parallels  that followed for the operators. The starting point
 is the 
association of a function  
  $I_{\Gamma}(\xi , \beta ) $  with each phase space function 
$\A(t,f)$ by the relation: 
\be
I_{\Gamma}(\xi , \beta ) \equiv 
\int_{\Gamma} {\cal{A}}(t,f) \delta ((t-\xi )f- \beta ) \; dtdf
\label{ip}
\ee
For a given $\xi $, the function defined by (\ref{ip}) is interpreted as 
 the integral of $\A(t,f)$ on the various orbits 
of the subgroup $G_{\xi}$.
The study of the properties of  function  $I_{\Gamma}(\xi , \beta ) $ 
follows the same steps as in Section 3.1. For practical reasons, the notation 
 $I_{\Gamma}(\A ;\xi , \beta ) $ will also be used for expression (\ref{ip}).
\begin{description}
\item
{\em $G_{\xi}$-invariance of $I_{\Gamma}(\xi , \beta ) $ } \par

The integral in (\ref{ip}) is clearly invariant under 
the action of $G_{\xi}$ on ${\cal{A}}(t,f)$ defined by (\ref{aeu}). 
\item
{\em Affine covariance}\par 
When $\A (t,f)$ is transformed
into $\A'(t,f) \equiv \A (e^ut+b, a^{-1}f)$, the function $I_{\Gamma}(\A; \xi , \beta )$
becomes:
\be
I_{\Gamma}(\A'; \xi , \beta )=I_{\Gamma}(\A; e^u\xi +b , \beta )
\label{invig}
\ee
This  relation connects the integrals of the function $\A'$ on the $G_{\xi}$ orbits 
to the integrals of $\A$ on the orbits of the subgroup $G_{\xi'}$, conjugate of $G_{\xi}$ 
by the transformation $(u,b)$ of the affine group.
\item
{\em Reconstruction of the function ${\cal{A}}(t,f)$} \par

Relation (\ref{ip}) defines in fact the Radon transform of 
${\cal{A}}(t,f)$ with respect to  arrays of hyperbolas parametrized by 
$\xi$ and $\beta$. To invert this 
transform, we introduce  the two-dimensional Fourier transform of $I_{\Gamma}(\xi , \beta )$ in a manner 
analogous to (\ref{ff}) and compute
\be
\widehat{I}_{\Gamma}(u,v) = 
\int_{\R} e^{-2i\pi ut } \,
{\cal{A}}(t, (u/v)) \; \frac{dt}{|v|}
\label{ia}
\ee
The inversion of this formula gives:
\be
{\cal{A}}(t,f)= f\int_{\R} e^{2i\pi vft} \, \widehat{I}_{\Gamma} (fv,v) |v| \; dv
\label{invp}
\ee

\item
{\em Scalar product of two symbols} \par
Consider two functions ${\cal{A}}(t,f)$, 
$ {\cal{B}}(t,f)$ and their corresponding 
$\widehat{I}_{\Gamma}$ functions. The use of relation (\ref{invp}) 
and Parseval's formula  yield:
\be
\int_{\Gamma} {\cal{A}}(t,f){\cal{B}}^*(t,f) \; dtdf = 
\int_{\R ^2} \widehat{I}_{\Gamma} (\A;u,v)
\widehat{I}_{\Gamma} ({\cal B}^*;u,v) \, |u| \; dudv
\label{u2}
\ee
\end{description}

\subsection{The geometric correspondence rule}
 So far, 
 operators on the Hilbert space $\cal{H}$  
 and functions on the 
phase space $\Gamma$ have been characterized respectively 
by  functions $I_{\cal{H}}(\xi ,\beta)$ or 
$I_{\Gamma}(\xi ,\beta)$. Moreover, for  each value of $\xi$, 
both functions $I_{\cal{H}}(\xi ,\beta )$ and $I_{\Gamma}(\xi ,\beta )$ are invariant by action 
of the subgroup $G_{\xi}$. Thus it is possible to define 
the correspondence between operators and symbols by 
requiring that
\be
I_{\cal{H}}(\xi ,\beta) \equiv 
I_{\Gamma}(\xi ,\beta ) 
\label{iii}
\ee
and to denote the common function by $  I(\xi ,\beta)$. 
This relation is stable by the affine group as seen from 
properties (\ref{invih}) and (\ref{invig}).
Actually, the consistency with the group action implies only the 
proportionality of the two members of (\ref{iii}). The interest of the 
strict identification which has 
been adopted is  to ensure that
the identity operator will be symbolized by  the function one. \par

According to  (\ref{hatih}) and (\ref{ia}),  the Fourier transform of 
relation (\ref{iii}) leads to:
\begin{eqnarray*}
\widehat{I}(u,v) & \! \! \! \! = \! \! \! \! &
A\left(\frac{ue^{(v/2)}}{2\sinh (v/2)}, 
\frac{ue^{(-v/2)}}{2\sinh (v/2)} \right) 
\left( \frac{u}{2\sinh (v/2)} \right) ^{2r} 
\frac{|u|}{4\sinh^2 (v/2)} \, Y(u/v)  \\
& \! \! \! \! = \! \! \! \! &
\int_{\R} e^{-2i\pi ut } \,
{\cal{A}}(t, (u/v)) \; \frac{dt}{|v|}
\end{eqnarray*}
The equality of the two right hand sides   gives the relation between the kernel $A(f_1,f_2)$ 
and the symbol $\A (t,f)$. This is the correspondence rule that can be 
rewritten either as a formula giving the operator kernel in terms of the symbol 
(Weyl mapping) or as a formula giving the symbol in terms of the kernel (Wigner correspondence).
The result is the following: 
{\result   
 The operator {\bf A}  corresponding to the function 
${\cal{A}}(t,f)$, defined on the time-frequency half-plane $\Gamma$, 
 is given by its kernel (defined in(\ref{kernel})) as:
\be
A(f_1,f_2) = (f_1f_2)^{-r-1} \frac{f_1-f_2}{\ln (f_1/f_2)} 
\int_{\R} e^{-2i\pi (f_1-f_2)t} \, {\cal{A}} \left( t,\frac{f_1-f_2}{\ln (f_1/f_2)} \right)
\; dt
\label{averop}
\ee
 
Conversely,  the  operator $\ab$ 
defined by its 
kernel $A (f_1,f_2)$, 
is represented on phase space 
by a symbol ${\cal{A}}(t,f)$    according to:
\be
{\cal{A}}(t,f) = f^{2r+2} \int_{\R} e^{2i\pi vft} \, 
A \left( \frac{fve^{v/2}}{2\sinh (v/2)}, \frac{fve^{-v/2}}{2\sinh (v/2)} \right) 
\left( \frac{v}{2\sinh (v/2)} \right)^{2r+2} \; dv
\label{opvera}
\ee
In addition, the following unitarity property holds:
\be
\int_{\Gamma} {\cal{A}}(t,f) {\cal{B}}^*(t,f) \; dtdf = 
\mbox{Tr} \, (\ab{\bb}^{\dagger})
\label{unitary}
\ee
\label{wwrule}
}
 Among the properties of the correspondence, it can be deduced from (\ref{opvera}) 
that the identity operator with kernel:
 \be
I(f_1,f_2) = (f_1f_2)^{-r-1/2} \, \delta(f_1-f_2) 
\ee
has a symbol equal to one, and that operators such that 
\be
A (f_1,f_2)= A^*(f_2,f_1)
\label{symmetry}
\ee
are in correspondence with real-valued functions $\A(t,f)$.\par
 At this point, it would be desirable 
to   characterize in a precise way 
classes of operators and functions that are put in correspondence 
by the above rule. Clearly, because of property (\ref{unitary}), 
 Hilbert-Schmidt operators can be associated with functions in the 
  space $L^2(\Rbig^+; f^{2r+1}df)$.  
In fact,  it is possible to characterize other classes of operators by their symbols 
  \cite{unterberger2}-\cite{arnal2}, as 
is done in the usual Weyl case with bounded operators \cite{bornes}-\cite{bornes2}. 
But this is outside the scope of the present paper.
\par
Result (\ref{wwrule}) concerns only functions defined on the positive half-line 
and their corresponding operators. Some side comments concerning its possible extension 
are given in the following remark.\par
{\remark \rm 
A direct extension of result (\ref{wwrule})  can be obtained if a function $\A (t,f)$ defined 
on the whole plane is substituted in (\ref{averop}) and if negative values for
$f_1$ and $f_2$ are allowed. 
However, the result will be a 
function $A (f_1,f_2)$ defined only in the first and third quadrant of the $(f_1,f_2)$ 
plane. 
The regions 
where $f_1$ and $f_2$ have different signs are definitely 
forbidden in the above symbolic calculus.
\label{32}}\par

\section{Some special cases}
\setcounter{equation}{0}

\subsection{Symbol of the projector on a ${\bos \beta}$-subspace of ${\cal H}$}\label{41}

The $\beta$-subspace associated with subgroup $G_{\xi}$ 
is characterized by the function $\psi^{\xi}_{\beta} (f)$ introduced 
in (\ref{psi}). The projector on this subspace is 
the linear operator defined by:
\be
S(f) \,  \longrightarrow \, (\psi^{\xi}_{\beta},S) \, \psi^{\xi}_{\beta}(f) 
\ee
\noindent 
 where the scalar product has the form (\ref{scalar}). The corresponding kernel  
    is given by (cf. (\ref{kernel}):
\be
\Pi^{\xi}_{\beta} (f_1,f_2) = \psi^{\xi}_{\beta}(f_1) \psi^{\xi *}_{\beta} (f_2)
\label{projb}
\ee
A direct application of formula (\ref{opvera}) shows that the symbols 
associated with (\ref{projb}) is the distribution 
$\delta((t-\xi )f -\beta)$ 
with support in $\Gamma$ localized on the $\beta$-orbit of subgroup $G_{\xi}$.

\subsection{Affine Wigner function as a particular symbol}
Each signal $S(f)$ defines a real linear functional on the set of 
hermitian operators by:
\be
<\ab > = (\ab S,S) = (S, \ab S)
\label{ln}
\ee
\noindent where the scalar product is given by (\ref{scalar}).
 \par
This functional can also be written as:
\be
<\ab > = \mbox {Tr} \, ({\bos \Pi}_S \, \ab )
\label{mean1}
\ee
\noindent where the trace operation has been defined by (\ref{trace}) 
and where ${\bos \Pi}_S$ is the projector on signal $S(f)$ with 
kernel given by:
\be
{ \Pi}_S(f_1,f_2) = S(f_1)S^*(f_2)
\label{kernelpi}
\ee
From the reciprocal correspondence   (\ref{averop}), (\ref{opvera})  and  its 
properties, it results that each real linear functional on the set 
of hermitian operators 
can be expressed as a real linear functional 
of the real-valued functions 
on the time-frequency half-plane. 
The   Wigner function for the signal $S(f)$ is defined as the function 
(or distribution) 
${\cal P}(t,f)$ on the time-frequency half-plane which allows to rewrite (\ref{mean1}) 
in the form:
\be
<\ab > = \int_{\Gamma} {\cal P}(t,f) \A(t,f) \; dtdf
\label{mean2}
\ee
\noindent where $\A(t,f)$ is the symbol  of operator $\ab$. 
From this definition, it is clear that the form of the Wigner function  
is fundamentally dependent 
of the symbolic calculus which is adopted.\par

The equality of the right-hand sides of (\ref{mean1}) and (\ref{mean2}) 
gives a relation of type (\ref{unitary}) (unitarity property) provided the Wigner function 
${\cal P}(t,f)$  is identified with the geometric symbol of the projector
  ${\bos \Pi}_S$. This makes it possible to obtain the relevant Wigner function by 
applying the correspondence rule (\ref{opvera}) to the kernel (\ref{kernelpi}).
The result is the affine Wigner function \cite{cr}:
\be
{\cal P}(t,f) = f^{2r+2} \int_{\R} e^{2i\pi vft} 
S\left( \frac{fve^{v/2}}{2\sinh (v/2)} \right) 
S^*\left( \frac{fve^{-v/2}}{2\sinh (v/2)} \right) 
\left( \frac{v}{2\sinh (v/2)} \right)^{2r+2} \; dv  
\label{wigner}
\ee
This function can be introduced in various ways. 
In the above development, it arises as a consequence of the general correspondence 
defined by  (\ref{averop}),(\ref{opvera}).
\par
Expression (\ref{wigner}) can also be obtained directly by writing the unitarity relation (\ref{unitary}) 
with one symbol equal to 
${\cal P}(t,f)$ and the other one equal to the localized distribution 
$\delta(t-t_0) \, \delta (f-f_0)$.
This leads to the relation:
\be
{\cal P}(t_0,f_0) = \mbox{Tr} \,( {\bos \Pi}_S {\bos \Delta}_{t_0,f_0}) 
\label{mean}
\ee
\noindent where ${\bos \Delta}_{t_0,f_0}$ is the operator with symbol 
$\delta(t-t_0) \, \delta (f-f_0)$. The kernel of this operator is given by 
(\ref{averop}) and reads:
\be
\Delta (f_1,f_2) = (f_1f_2)^{-r-1} \frac{f_1-f_2}{\ln f_1/f_2} 
 e^{-2i\pi(f_1-f_2)t_0} \,  
\delta \left( \frac{f_1-f_2}{\ln f_1/f_2} -f_0 \right)  
\ee
Function $\Pc (t,f)$ can then be obtained by (\ref{mean}).
  These results are summarized by:
{\result  
The   Wigner function of signal $S(f)$ associated with the geometric correspondence is given 
by (\ref{wigner}). It  
can be obtained either as the symbol of the 
projector ${\bos \Pi}_S$ on signal $S(f)$, or as the mean value (\ref{mean}) of the operator 
${\bos \Delta}_{t_0,f_0}$ whose symbol is $\delta (t-t_0) \delta(f-f_0)$.
}
\par
\vspace{3mm}

The unitarity property (\ref{unitary}) gives  directly the formulas:
\be
\int_{\Gamma} \Pc(t,f) \; dtdf \,  = \; \parallel S \parallel^2
\ee
\begin{eqnarray*}
\int_{\Gamma} \Pc_1(t,f)\Pc_2(t,f) \; dtdf & = & \mbox{Tr} \; ({\bos \Pi}_{S_1}{\bos \Pi}_{S_2})
\\[1mm]
& = & |(S_1,S_2)|^2
\end{eqnarray*}
 \noindent where 
${\Pc}_1$ and ${\Pc}_2$ stand for  the Wigner functions corresponding to 
 $S_1$ and $S_2$ respectively.\par
Many other properties could be mentioned which are related to the interpretation 
of the Wigner function as a time-frequency representation \cite{boudreaux}, \cite{flandrin}.\par
\subsection{Exponentials of generators and Weyl's formulation of the correspondence}
We will now show that the 
  symbol of the operator $\eb_{u_0v_0} \equiv e^{-2i\pi (u_0 \bos{\beta} +v_0 \bos{f})}$,  
where  $\bos{\beta}$ and $\bos{f}$ are the generators defined by 
(\ref{genb}) and (\ref{genf}), is given by the function ${\cal E}_{u_0v_0}(t,f) = 
e^{-2i\pi (u _0 tf +v_0 f)} $.\par
 
It follows from the 
developments of Section 2.1 that the operator ${\eb}_{u_0v_0}$ satisfies 
the following eigenvalue equation:
\be
e^{-2i\pi (u_0 \bos{\beta} +v_0 \bos{f})}
 \psi_{\beta }^{-v_0/u_0} (f) = e^{-2i\pi \beta u_0} 
\psi_{\beta }^{-v_0/u_0} (f)
\label{esurs}
\ee
The kernel of $\eb_{u_0v_0}$ is obtained directly from the spectral 
decomposition: 
\be
{ E}_{u_0v_0}(f_1,f_2) = \int_{\R} e^{-2i\pi \beta u_0} \, 
\psi_{\beta }^{-v_0/u_0} (f_1) \psi_{\beta }^{*-v_0/u_0} (f_2)
\; d\beta
\label{em}
\ee
 and has the explicit expression:
\be
E_{u_0v_0}(f_1,f_2) = e^{(r+1)u_0} \, \exp \{ -2i\pi f_1 (v_0/u_0)(e^{u_0}-1)\} \,
 \delta (f_1e^{u_0}-f_2) \, f_2^{-2r-1}
\ee

Substituting this result in (\ref{opvera}), we can write  the symbol of 
${\eb}_{u_0v_0}$ as:
\begin{eqnarray}
{\cal E}_{u_0v_0}(t,f) &  = &  \int_{\R} e^{2i\pi vft} \, e^{(r+1)u_0} 
\exp \{ -2i\pi (f(v_0/u_0)(e^{u_0}-1)\lambda(v)) \} \nonumber \\[2mm]
& &  \quad \times  \; 
\lambda(-v)\, e^{(2r+2)(v/2)}   \, 
\delta (e^{u_0}\lambda (v)- \lambda (-v))\; dv
\label{inte}
\end{eqnarray}
where, by definition 
\be
\lambda (v) = \frac{ve^{v/2}}{2\sinh (v/2)}
\label{lambda}
\ee
Since
\be
\delta (e^{u_0} \, \lambda (v) - \lambda (-v)) 
= \frac{1}{\lambda (u_0)} \, \delta (v+u_0) 
\ee
the result of integration in (\ref{inte}) is just:
\be
{\cal E}_{u_0v_0}(t,f) = e^{-2i\pi (u_0ft+v_0f)}
\ee

Thus, we have established the correspondence
\be
{\cal E}_{u_0v_0}(t,f) \equiv e^{-2i\pi (u_0ft+v_0f)} \quad 
\longleftrightarrow \quad 
\eb_{u_0v_0} = e^{-2i\pi (u_0 \bos{\beta} +v_0\bos{f})}
\label{exprule}
\ee
Remark that the variable $\beta = tf$ rather than $t$ appears 
naturally. It will sometimes be convenient to use it by performing the change of 
variables according to:
\be 
\tilde{\cal{A}} (\beta,f)  =  \left[{\cal A}(t,f) \right]_{t=\beta /f}
\label{coru2}
\ee
  \par
 
The correspondence (\ref{exprule}) allows to write the operator $\ab$  
  in terms of its symbol $\A(t,f)$ by  the Weyl formula:
\be
\ab=\int 
\widehat{\cal{A}} (u ,v) \, 
e^{-2i\pi (u \bos{\beta} +v \bos{f})} \; dudv 
\label{therule}
\ee
where  $\widehat{\cal {A}} (u ,v)$ is the Fourier transform
of $\tilde{\cal{A}} (\beta,f)$ given by:
\be
\widehat{\cal {A}} (u ,v)  \equiv \int_{\R \times \R^+}  
e^{2i\pi (u  \beta +vf)} \, 
\tilde{\A} (\beta,f)   \;  d\beta df
\label{af}
\ee

The correspondence between functions and operators    defined by 
(\ref{therule}) and (\ref{af}) has a form analogous to Weyl's original rule. In fact, it is 
what L.Cohen calls Weyls'rule for operators $\bos{\beta}$ and $ \bos{f}$  
\cite{cohen}. 
It can also be obtained as a by-product of Kirillov's analysis \cite{kirillov} 
as 
shown by A.Unterberger \cite{unterberger2} (see also R.Shenoy and T.Parks \cite{shenoy}).
\par

{\remark \rm
The symbols of the self-adjoint operators ${\bos \beta}$ and ${\bos f}$ 
are immediately seen to be $\beta =tf$ and $f$ respectively.
The time operator, which is especially important in signal analysis, 
can be defined as the operator with symbol $t$. A direct computation 
leads to the result:
\be
{\bos t} = (1/2) \, ({\bos f}^{-1} {\bos \beta} + 
{\bos \beta} {\bos f}^{-1})
\ee
 or:
\be
{\bos t} =- \frac{1}{2i\pi } \left( \frac{r+1/2}{ f} + \frac{d}{df} 
\right)
\ee
Such an operator is not essentially self-adjoint  on the 
Hilbert space $L^2(\Rbig^+; f^{2r+1}df)$ and hence has no conventional spectral 
decomposition. However, it is known  from the deformation quantization theory on 
phase space \cite{flato} that such a difficulty can be easily overcome 
  and does not affect the physical interpretation.}
\section{Star product on the symbols}
\setcounter{equation}{0}

Operators on the Hilbert space ${\cal H}$ have been mapped to functions on the 
time-frequency half-plane $\Gamma$ by the rule (\ref{opvera}).  
Since the algebra of operators is non-abelian,  the product of operators   is mapped 
into
a composition law of functions that  cannot be the ordinary product. 
This situation is familiar from  Weyl's calculus where the composition law 
of functions is known as Moyal's product and is invariant by Heisenberg's group.  
  Analogous products, invariant by different groups, have been 
constructed by deformation of the usual product and are generally referred to as star 
products ($\star $-products) \cite{flato}. In the present case, 
the composition law of symbols corresponding to the product of operators leads directly to a 
star product that is invariant by the affine group.
\par
In the following, it will be convenient to use another 
 notation for the operator \\
$ \exp \{ - 2i\pi (u \bos{\beta} +v \bos{f}) \}$. In fact, according to 
(\ref{esurs}) and (\ref{em}), the action of that operator on a signal $S(f)$ can be written as
\be
 \exp \{ - 2i\pi (u \bos{\beta} +v \bos{f}) \} S(f) = 
e^{(r+1)u} \, \exp \{ -2i \pi f(v/u)(e^u-1) \} S(fe^u)
\label{rewr}
\ee
Comparing this expression with the definition (\ref{uab}) of  representation $\ub$, we 
are able to rewrite (\ref{rewr}) as
\footnote{ This result   can also be obtained directly by noticing that,  
in the description of the affine group by  
$2 \times 2$ matrices, 
the exponentiation of the Lie algebra element 
defined by $(u,v)$ is equal to:
\[
\exp 
\begin{array}{|cc|}
u & v \\
0 & 0 
\end{array}
=
\begin{array}{|cc|}
e^u & (v/u)(e^u-1) \\
0 & 1 
\end{array}
\]
}:
\be
 \exp \{  -2i\pi (u \bos{\beta} +v \bos{f}) \} \, 
S(f) = \ub \left(u,\frac{v}{u} (e^u-1) \right) \, S(f) 
\label{equalu}
\ee
The use of representation $\ub (u,b)$  allows to take advantage of the group 
law and to write  the composition of operators as:
\be
\ub(u,b) \ub(u',b') = \ub(u+u',b+e^u \, b')
\label{glaw}
\ee
\noindent It shows directly that the product of two operators of type (\ref{rewr}) 
is an operator of the same type.
With notation (\ref{equalu}), the form  (\ref{therule}) of the correspondence rule 
(\ref{averop}) between functions and operators can  be written as:
\be
\ab = \int_{\R^2} \widehat{\A}(u,v) \,\ub
\left (u, v\frac{e^u-1}{u} \right) \; dudv
\label{coru}
\ee
\noindent where $\hat{\cal {A}} (u ,v)$ is defined in (\ref{af}).
\par 
Let $\ab$ and $\bb$ be operators given in terms of their symbols $\tilde{\A}(\beta ,f)$ 
and $\tilde{\B}(\beta ,f)$ 
 by rule  (\ref{coru}),(\ref{af}). Their product 
is equal to:
\begin{eqnarray*}
\ab{\bb} & = &  
 \int_{\R^4} \widehat{\A}(u_1,v_1) \widehat{\B}(u_2,v_2) 
\ub \left ({u_1}, v_1\frac{e^{u_1}-1}{u_1} \right) \, 
\ub \left ({u_2}, v_2\frac{e^{u_2}-1}{u_2} \right)
 \; du_1dv_1du_2dv_2 \\[2mm]
 \mbox{or, using} & \! \! \mbox{ (\ref{glaw})} \! \! &: \\[2mm]
&  = & \int_{\R^4} \widehat{\A}(u_1,v_1) \widehat{\B}(u_2,v_2) 
\ub \left ({u_1+u_2}, v_1\frac{e^{u_1}-1}{u_1} + e^{u_1}v_2 
\frac{e^{u_2}-1}{u_2} \right) \; du_1dv_1du_2dv_2
\end{eqnarray*}
This product   can be rewritten in the form (\ref{coru}) of the correspondence rule 
if variables $(u,v)$ are introduced such that:
\be
u=u_1+u_2, \quad 
v \frac{e^u-1}{u} = v_1 \frac{e^{u_1}-1}{u_1} +e^{u_1} \, v_2 \frac{e^{u_2}-1}{u_2} 
\ee
The result is:
\be
\ab \bb =  \int_{\R^2} \widehat{\C}(u,v) \,\ub
\left (u, v\frac{e^u-1}{u} \right) \; dudv
\label{aabb}
\ee
with $\widehat{\cal C}(u,v)$  defined by:
\begin{eqnarray}
\widehat{\cal C}(u,v) & = & \int_{\R^2 \times \R^2} 
\widehat{\A}(u_1,v_1) \widehat{\B}(u_2,v_2) 
\delta (u-u_1-u_2) \label{cuv}  \\[2mm] 
& & \times \, \delta \left(v-\left(  v_1\frac{e^{u_1}-1}{u_1} + e^{u_1}v_2 
\frac{e^{u_2}-1}{u_2} \right) \frac{u_1+u_2}{e^{u_1+u_2}-1} \right) 
\; du_1dv_1du_2dv_2
\nonumber
\end{eqnarray}
It follows from (\ref{aabb}) that $\widehat{\cal C}(u,v)$ is the Fourier transform of the 
symbol of the operator $\ab \bb$.

To obtain an explicit expression of the star product 
$\tilde{\A} (\beta ,f) \star \tilde{\B} (\beta ,f)$, it suffices now to invert 
  the Fourier transform of $\widehat{\cal C}(u,v)$  and to express the result 
   in terms of 
$\tilde{\A} (\beta ,f)$ and $\tilde{\B} (\beta ,f)$. 
In a first step, we obtain:
\begin{eqnarray}
\lefteqn{ \tilde{\A}(\beta ,f) \star \tilde{\B} (\beta ,f)=} \label{starprod} \\[3mm]
& {\displaystyle \int} \widehat{\A}(u_1,v_1) \widehat{\B}(u_2,v_2)
e^{2i\pi ((u_1+u_2)\beta + (v_1+v_2)f)} 
e^{ -2i\pi f h(u_1,v_1,u_2,v_2) }\; du_1dv_1du_2dv_2 
\nonumber
\end{eqnarray}
\noindent where the function $h$ is defined by:
\be
 h(u_1,v_1,u_2,v_2)  \equiv   
  (v_1+v_2) 
-\frac{u_1+u_2}{e^{u_1+u_2}-1} \left(  v_1\frac{e^{u_1}-1}{u_1} + e^{u_1} v_2 
\frac{e^{u_2}-1}{u_2} \right)  \nonumber 
\ee \noindent or, after some manipulations:
 \be
h(u_1,v_1,u_2,v_2)
= (v_1u_2-v_2u_1) \frac{u_1e^{u_1}(e^{u_2}-1) -u_2(e^{u_1}-1)}{u_1u_2(e^{u_1+u_2}-1)}
\label{belex}
\ee
 The exponential of function 
$h(u_1,v_1,u_2,v_2)$ in (\ref{starprod}) is what 
makes the difference with a classical product.
It can be replaced by its 
series expansion so that the integrals in (\ref{starprod}) can be performed.  In 
the process, the following replacements  take place:
\be
v_i \longrightarrow \frac{1}{2i \pi} \partial _{f_i}, \quad 
u_i \longrightarrow \frac{1}{2i \pi} \partial _{\beta_i}, \quad i=1,2
\ee
The development of the star product can thus be written as:
\begin{eqnarray}
\lefteqn{\tilde{\A}(\beta ,f) \star \tilde{\B} (\beta ,f) = \tilde{\A}(\beta ,f)  \tilde{\B} (\beta ,f)} 
\label{thedev} \\[2mm]
 &  + \, 
{\displaystyle \sum_{n=1}^{\infty}} {\displaystyle \frac{1}{ n!}}
\left( {\displaystyle \frac{f}{4i\pi \,}} (\partial_{f_2} \partial_{\beta_1}- 
\partial_{f_1} \partial_{\beta_2})  T(\partial_{\beta_1}, \partial_{\beta_2}) \right)^n
\tilde{\A}(\beta_1 ,f_1) \tilde{\B} (\beta_2 ,f_2)  
\left. \right| _{\stackrel{f_1=f_2=f}{ \beta_1=\beta_2= \beta} } 
\nonumber 
\end{eqnarray}
\noindent where the contribution of operator $T$ will be discussed below.\par
  If    (\ref{thedev}) is rewritten in terms of $t$ rather than $\beta$,
 using the definition:
\be
\beta = tf, \quad \partial_{\beta} = (1/f) \, \partial_t
\label{betat}
\ee
\noindent the result is reminiscent of the Weyl product. Actually, the difference 
lies in the presence of the operator 
$T(\partial_{\beta_1}, \partial_{\beta_2})$ which comes from the development in 
powers of $u_1$ and $u_2$  of:
\be
T(u_1,u_2) = 2 \frac{u_1e^{u_1}(e^{u_2}-1) -u_2(e^{u_1}-1)}{u_1u_2(e^{u_1+u_2}-1)}
\ee
 Operator $T$ introduces derivatives of all orders with respect to 
$\beta_1$ and $\beta_2$. At the lowest orders, one has
\begin{eqnarray}
\lefteqn{T(\partial_{\beta_1}, \partial_{\beta_2})=} \\          
& & 1+\frac{i}{12\pi}(\partial_{\beta_2}- \partial_{\beta_1}) 
+ \frac{1}{12(2\pi)^2}\partial_{\beta_1} \partial_{\beta_2}   
+ \frac{i}{720(2\pi)^3}(\partial_{\beta_2} - \partial_{\beta_1})
(\partial_{\beta_1}^2 + \partial_{\beta_2}^2+5\partial_{\beta_1}\partial_{\beta_2}) + 
\cdots \nonumber
\end{eqnarray}
As in    Weyl's case, the first term in (\ref{thedev}) 
is the usual product and the following terms 
introduce  derivatives of the symbols. It can be observed that the star product of two real-valued functions is not 
real. This is a direct  manifestation of the fact that the product of hermitian operators 
is not hermitian.\par
{\remark \rm 
From Remark \ref{32}, it follows that in a product of operators, 
the positive and negative frequency parts are multiplied separately. Hence,  the 
same property will hold for 
the star product if its expression   is extended to symbols defined on the whole plane. }
\par
\vspace{2mm}
The star product simplifies when one of the symbols depends only on one variable. 
This will be illustrated by computing  its expression  when $\tilde{\A}(\beta ,f)$ is 
equal either to $\beta$ or to a function of $f$ alone and when $\tilde{\B}(\beta ,f)$ is 
  the generic distribution:
\be 
 \Delta_{\beta_0,f_0} (\beta,f) = 
 \delta (\beta - \beta_0) \delta (f-f_0)
\label{setb}
\ee
 with Fourier transform   (\ref{af})  given by:
\be
\widehat{\Delta}_{\beta_0,f_0}(u,v)   =     \, e^{-2i\pi (u\beta_0 +vf_0)}
\label{fourb}
\ee
\begin{description}
\item
{\em (i) Case where $\tilde{\A}(\beta,f)=\beta$} \par
The Fourier transform of $ \beta$ is  :
\be
\widehat{\A}(u,v)  =   -\frac{1}{2i\pi} \delta'(u) \, \delta(v)
\ee
 Computation of the star product   from (\ref{starprod}) and (\ref{belex})  gives:
\begin{eqnarray}
\lefteqn{\beta  \star \delta(\beta-\beta_0 )\delta(f-f_0)  = } \label{bstarb0} \\[3mm]
& - \frac{1}{2i\pi } {\displaystyle \int } \delta'(u_1) \, 
e^{-2i\pi [u_2(\beta_0 - \beta )+v_2f_0]} 
\, \exp \{2i\pi [u_1\beta + fv_2\frac{\lambda(u_1+u_2)}{\lambda(u_2)}] \} 
\; du_1du_2dv_2 & \nonumber
\end{eqnarray}
\noindent where $\lambda (u)$,  defined by (\ref{lambda}),  is such that:
\be
\lambda (-u) = e^{-u} \lambda (u)
\ee
Performing the $u_1$-integration, we can write (\ref{bstarb0}) as:
\begin{eqnarray}
 \beta  \star \delta(\beta-\beta_0 )\delta(f-f_0) & \! \! \! \! = \! \! \! \!  & 
    \beta \delta (\beta -\beta_0) \delta (f-f_0)  \label{bstarb} \\[3mm]
 &  \! \! \! \! +  \! \! \! \! &  f\int_{\R^2} e^{2i\pi [u_2(\beta -\beta_0) +v_2(f-f_0)]} 
\left(   \frac{d}{du_2} (\ln (\lambda ( u_2))) \right) v_2 \; du_2dv_2  
\nonumber
\end{eqnarray}

\item
{\em (ii) Case where $\tilde{\A}(\beta,f)$ depends only on $f$. }\par
The Fourier transform of $\tilde{\A}(\beta,f)= g(f)$, with $g(f)$ an arbitrary function, 
 is equal to:
\be
\hat{\A}(u,v) = \delta (u )\, \int_{\R} g(f) e^{-2i\pi v f}   \; df
\label{fourg}
\ee
 Substitution of the  Fourier transforms  (\ref{fourb}) and (\ref{fourg}) in the 
expression of the star product  (\ref{starprod})-(\ref{belex}) gives: 
 \be
 g(f) \star \delta(\beta-\beta_0 )\delta(f-f_0)  =
  \int_{\R^2} g(f\lambda (-u)) e^{2i\pi [u(\beta -\beta_0) +v(f-f_0)]}\; dudv
\label{fstarb}
\ee

\end{description}

\section{Star bracket of symbols and Hamiltonian flow}
\subsection{Definition of the star bracket}
A Lie bracket is now defined on operators by the operation:
\be
-2\pi i [\ab, \bb] = -2\pi i (\ab \bb - \bb \ab)
\label{mqbra}
\ee
This bracket defines a composition law on the subspace of hermitian operators. 
When transformed by the correspondence rule, using the expression for the star product, this bracket 
gives rise to an operation on real-valued functions that will be called star bracket 
and denoted by $\{ \; , \; \}_{\star}$.
Thus the star bracket of symbols $\tilde{\A} (\beta ,f)$ and $\tilde{\B} (\beta ,f)$ is 
defined by:
 \be
\{ \tilde{\A} ,\tilde{ \B} \}_{\star}(\beta ,f) = - 
2\pi i (\tilde{\A}(\beta ,f) \star \tilde{\B}(\beta ,f) -
\tilde{\B}(\beta ,f) \star \tilde{\A}(\beta ,f))
\label{brac}
\ee
 According to (\ref{thedev}), the bracket can be written as;
\begin{eqnarray}
& \{ \tilde{\A}(\beta ,f) , \tilde{\B}(\beta ,f) \}_{\star}     =   \left[
f(\partial_{f_1}\partial_{\beta_2}-\partial_{f_2}\partial_{\beta_1}) (1+
\frac{1}{12(2\pi)^2}\partial_{\beta_1} \partial_{\beta_2}  +\cdots) \right. & \label{poidev}
\\[3mm]
& +  \left.{\displaystyle \sum}_{n=2}^{\infty} {\displaystyle \frac{2i\pi}{ n!}}
\left( {\displaystyle \frac{f}{4i\pi \,}} (\partial_{f_1} \partial_{\beta_2}- 
\partial_{f_2} \partial_{\beta_1}) \right)^n [T^n(\partial_{\beta_2}, \partial_{\beta_1}) 
- (-1)^n T^n(\partial_{\beta_1}, \partial_{\beta_2})] \right] \nonumber \\[3mm]
 &  \quad \quad \times \tilde{\A}(\beta_1 ,f_1) \tilde{\B} (\beta_2 ,f_2)  
\left. \right| _{\stackrel{f_1=f_2=f}{ \beta_1=\beta_2= \beta} } & \nonumber
\end{eqnarray}

\par
The first term  in $f$ is a Poisson 
bracket given by: 
\be
\{ \tilde{\A} (\beta ,f), \tilde{\B} (\beta ,f)\}_P = f \left(
{\displaystyle \frac{\partial \tilde{\A}(\beta ,f)}{\partial{\beta}} } 
{\displaystyle \frac{ \partial \tilde{\B} (\beta ,f)}{\partial f}} -
{\displaystyle \frac{\partial \tilde{\A} (\beta ,f)}{\partial {f}}}
{\displaystyle \frac{  \partial \tilde{\B} (\beta ,f)}{\partial {\beta} }} \right)
\label{poisson}
\ee

 In fact, if the variable $\beta $ is replaced by its definition (\ref{betat}), 
the expression (\ref{poisson})   becomes identical with the usual Poisson bracket on 
$\Gamma$. This is the same bracket as   obtained by Kirillov's theory \cite{kirillov} 
when considering 
phase space as an orbit of the coadjoint representation.\par
 
The form (\ref{poidev}) of the star bracket  
 is reminiscent of that obtained with the Weyl calculus.
However, due to the fact that operator $T(\partial_{\beta_1},\partial_{\beta_2})$ 
has no symmetry, derivatives of arbitrary order will in general 
appear in the expression of the bracket.
\par
Some  symbols are such that their star bracket with any other function    has the form  
of a Poisson bracket. 
In deformation quantization,  such symbols  
 are called  {\em preferred observables}                    
 \cite{flato}, \cite{arnal}, \cite{daste}. A typical example is the harmonic 
oscillator Hamiltonian on $\Rbig^2$ with Weyl quantization.
\par
 In the present case, 
the preferred observable are represented by the symbols  $\beta $, $f$ and $\ln f$. 
Explicitly, their star 
brackets with any symbol 
  ${\cal X}(\beta ,f)$ are given by: 
\begin{eqnarray}
\{ \beta , {\cal X}(\beta ,f) \}_{\star} & = & \, f \,   
\frac{\partial}{\partial  f}
 {\cal X}(\beta ,f) \label{poib} \\[2mm]
\{f,{\cal X}(\beta ,f)  \}_{\star} & = & - \, f \,  
\frac{\partial}{\partial \beta}
  {\cal X}(\beta ,f) \label{poif} \\[2mm]
\{\ln f,{\cal X}(\beta ,f)  \}_{\star} & = & -\,   
 \frac{\partial}{\partial \beta}{\cal X}(\beta ,f)
   \label{poilnf}
\end{eqnarray}
The derivation of these formulas can be based on expressions 
(\ref{bstarb}), (\ref{fstarb})  and on their counterpart with reverse order in the 
star products. Final results follow from the relations:
\be
\lambda (u) -\lambda (-u) = u, \quad \lambda (-u) = e^{-u} \, \lambda (u)
\ee
which are verified by the function (\ref{lambda})

\subsection{Extended covariance and Hamiltonian flows}

Consider a one-parameter group of transformations on signals $S(f)$ defined by 
an operator $\ob$ such that:
\be
S(f;\alpha ) = e^{-2i\pi \alpha \ob} \, S(f;0) 
 \label{flow}
\ee
\noindent with $S(f;0)=S_0(f)$ given. 
  Signal $S(f;\alpha)$ verifies the equation:
\be
\displaystyle \frac{\partial S(f;\alpha )}{\partial \alpha} = -2i\pi \alpha \,  \ob \, S(f;\alpha )
\ee
and the projector ${\bos \Pi}$ on $S(f; \alpha)$ verifies:
\be
{\displaystyle \frac{\partial {\bos \Pi} }{\partial \alpha}} =
 - 2 i \pi [ \, \ob , {\bos \Pi}  \, ]
\label{von}
\ee
Now, going to phase space, we know that the geometric symbol of ${\bos \Pi}$ is the 
affine Wigner function   (\ref{wigner}) for signal $S(f;\alpha)$, and that the 
symbol of the bracket $-2i\pi [\; ,\; ]$ is the star bracket $\{ \; ,\; \}_{\star}$.
Thus equation (\ref{von})   becomes the following equation in $\Gamma$:
\be
{\displaystyle \frac{\partial {\cal P}(t,f;\alpha )}{\partial \alpha}} = 
\{ {\cal O}(t,f) \; , \;  {\cal P}(t,f;\alpha ) \}_{\star}
\label{hamilton}
\ee
\noindent where ${\cal O}(t,f)$ is the symbol of operator $\ob$.\par
 Consider now the case where the symbol ${\cal O}(t,f)$ has  the particular form:
\be
{\cal O}_H (t ,f) = \mu tf + \nu f + \sigma \ln f
\label{Hamiltonian}
\ee
\noindent where $\mu$, $\nu$ and $\sigma$ are real constants. 
From the values of the  
 star brackets (\ref{poib}), (\ref{poif}) and (\ref{poilnf}), it follows that 
  the bracket in (\ref{hamilton}) reduces to Poisson's bracket.  As a result, 
equation (\ref{hamilton})  takes the form of a Liouville equation with 
Hamiltonian equal to ${\cal O}_H(t,f)$:
\begin{eqnarray}
{\displaystyle \frac{\partial {\cal P}(t,f;\alpha )}{\partial \alpha}} & = & 
\{  \mu tf + \nu f + \sigma \ln f, {\cal P}(t,f;\alpha ) \}_P \label{liou}
\\[2mm]
& = & \mu f  {\displaystyle \frac{\partial \Pc (t,f;\alpha )}{\partial f}}-
(\mu t + \nu + (\sigma /f))
{\displaystyle \frac{\partial \Pc (t,f;\alpha )}{\partial t}}
 \nonumber 
\end{eqnarray}
 This means that the $\alpha$-evolution of the Wigner function is identical to 
that of an  incompressible fluid in 
the time-frequency half-plane. 
Integration of   equation (\ref{liou}) with the 
initial condition 
\be
\Pc (t,f;0) = \Pc_0(t,f)
\ee
 gives:
\be
\Pc (t,f;\alpha)= \Pc_0 
\left( ( - \nu \mu^{-1} + (t+ \nu \mu^{-1}) \, e^{-\mu \alpha}   
- \sigma f^{-1} \,  \alpha ) \; , \; f \, e^{\mu \alpha}  \right)
\label{solgo}
\ee
The evolution of the Wigner function  in terms of parameter $\alpha$ has a 
counterpart in   Hilbert space  that 
 can be obtained 
using the   symbolic calculus. The operator $\ob_H$  for this evolution
is found  directly from (\ref{therule})-(\ref{af}) and has the expression:
\be
\ob_H = \mu {\bos \beta} + \nu \f + \sigma \ln \f
\label{geno}
\ee
\noindent where the operators ${\bos \beta}$ and $\f$  have been introduced in 
(\ref{genb}) and (\ref{genf}). 
The expression (\ref{flow}) written for the operator $\ob_H$ then gives the   
$\alpha$-evolution in the Hilbert space which corresponds to the phase space 
evolution (\ref{solgo}). In fact, this leads to a special form of  a general 
transformation which can be written as:
\be
(u,b,c):S(f) \longrightarrow  S'(f) = 
e^{(r+1)u} \, e^{-2i\pi bf} \, f^{-2i\pi c} \, S(e^uf), \quad u,b,c \in \Rbig
\label{uabc}
\ee
This transformation  is a projective unitary representation of a three-parameter group 
$G_0$ which has been studied in \cite{jmp}. It has been 
proved that transformations (\ref{uabc}) correspond to symplectic transformations 
in the phase space. This extended covariance of the calculus is at the basis of 
a generalization of the construction process that will now be considered.
 \par
 
\section{Other symbolizations with geometrical features.}

The method presented above to construct a symbolic calculus on the 
affine group can be 
embedded in a more general approach based on the systematic 
consideration of groups containing the affine group and 
having the time-frequency half-plane as coadjoint orbit.
Such groups  belong to the class  of  
  simply connected Lie groups 
having a coadjoint orbit of 
dimension smaller or equal to two \cite{cahen}. In this class, the groups     
  of interest are  
$SL(2,\Rbig )$ and the three-parameter solvable groups with   Lie algebras 
defined by  generators $X_a$, $X_b$ and $X_c$  satisfying
\cite{group}:
\begin{eqnarray}
& [X_a, X_b] = X_b,& \quad [X_a, X_c] = k \, X_c, \quad k\in \Rbig 
\label{lie1} \\[3mm]
& [X_a, X_b] = X_b,& \quad [X_a, X_c] = X_c + X_b 
\label{lie2}
\end{eqnarray}
   Examples of 
  symbolic calculi invariant by $SL(2,\Rbig )$ have been thoroughly studied  in
\cite{fronsdal}-\cite{ajunter}. Here, we will treat the cases of  
groups $G_k$ with Lie algebras defined by relations (\ref{lie1})  and (\ref{lie2})
\footnote{The notation $G_1$ will designate the group with Lie algebra defined in 
 (\ref{lie2}).} .
They provide a natural generalization of 
 the group $G_0$ considered so far in this paper and will be associated with 
new forms of symbolic calculus. 
All  groups $G_k$   have been introduced   
   in \cite{jmp}  as a tool to find explicit 
forms of 
time-frequency pseudo-distributions.  
    We refer to that  paper   for the details concerning 
the precise construction of their unitary representations (projective or not) 
according to the 
values of $k$.  Here we need only recall  the results   corresponding to
values of $k$ different from $0$ and $1$.
\par
  The groups $G_k$, $k\neq 0,1$ consist of elements $(u,b,c)$ with the 
composition law:
\be   
gg' =  (u+u', b+e^u \, b', c+e^{ku} \, c') 
\ee
Their  unitary representations in the Hilbert space ${\cal H}$ 
defined in Section \ref{21} are given by: 
 \be
\ub_k (u,b,c) S(f) = e^{(r+1)u} \, e^{-2i\pi (bf+cf^k)} \, S(e^u\,f)
\label{repgk}
\ee
The action of the groups in the time-frequency half-plane $\Gamma$ is:
\be
(t,f) \longrightarrow (e^u \, t+b +kce^{-(k-1)u} \, f^{k-1}, \, e^{-u} \, f)
\label{acgk}
\ee
\par
It is clear from (\ref{repgk}) and (\ref{acgk}) that the cases $k=0$ and $k=1$ 
 are  degenerate and deserve a special treatment 
(see \cite{jmp}).
\par
Following the same pattern as in section 2, we  determine the   subgroups 
of $G_k$ that are conjugate to the dilation subgroup. They are labelled
by two real parameters $\xi ,\eta$ and consist of:
\be
G_{\xi \eta} = \{ (e^u, \xi (1-e^u), \eta (1-e^{ku}) \}
\label{gxieta}
\ee
The restriction of representations (\ref{repgk}) to subgroups $G_{\xi \eta}$ is:
\be
\ub_{\xi \eta} S(f) = e^{(r+1)u} \, e^{-2i\pi (\xi(1-e^u)f+\eta (1-e^{ku})f^k)}S(f\, e^u)
\ee
The eigenfunctions of $\ub _{\xi \eta}$ are:
\be
\psi_{\beta}^{\xi \eta}(f) = f^{-2i\pi \beta -r-1} \, e^{-2i\pi (\xi f + \eta f^k)}
\ee
They transform under the full group $G_k$ as:
\be
\ub(u,b,c) \psi_{\beta}^{\xi \eta} (f) = e^{-2i\pi \beta u} \, 
\psi_{\beta}^{\xi e^u +b , \eta e^{ku} +c}(f)
\label{full}
\ee 
 In phase space $\Gamma$, the curves invariant by action (\ref{acgk}) restricted 
to $G_{\xi \eta}$ are given by:
\be
(t-\xi )f -k \eta f^k = \tilde{\beta}
\label{curvegk}
\ee
A local study of  $\psi_{\beta}^{\xi \eta}(f)$ analogous to that performed in Section \ref{23} 
leads to 
the identification:
\be
\tilde{\beta} = \beta
\ee
As in Section 2, we have achieved a correspondence between invariant structures in 
${\cal H}$ and $\Gamma$. But we now have one extra parameter labelling the vectors 
$\psi_{\beta}^{\xi \eta}(f)$ and the curves (\ref{curvegk}). To proceed with the construction, 
we must fix one of these parameters in such a way that the affine covariance of the 
procedure is preserved. 
Because of the transformation law (\ref{full}), the only possible choices  are either 
$\eta =0$ or $\beta = \beta_0$ fixed.
The case $\eta=0$ yields the calculus just developed in the preceding sections. The only 
possibility to obtain a new calculus is to fix the value of $\beta $ and to let $\eta$ free.
 \par
The same geometric construction as in Section 3 will be carried out. 
Consider an operator $\ab$ on ${\cal H}$ and form its diagonal matrix elements on 
$\psi_{\beta_0}^{\xi \eta}(f)$:
\be
I_{\cal H}^{\beta_0} (\xi , \eta) = (\psi_{\beta_0}^{\xi \eta} , \ab \psi_{\beta_0}^{\xi \eta} )
\label{diaggk}
\ee
Consider a function $\A(t,f)$ and form its integral with respect to curves (\ref{curvegk}):
\be
I_{\Gamma}^{\beta_0} (\xi ,\eta) = \int_{\Gamma} \A(t,f) \delta((t-\xi )f -k\eta f^k-\beta_0) \; dtdf
\label{radongk}
\ee
The study of the properties of functions $I_{\cal H}^{\beta_0} (\xi , \eta)$ 
and $I_{\Gamma}^{\beta_0} (\xi , \eta)$ follow the same steps as in Section 3.
\begin{description}
\item
{\em $G_{\xi \eta}$-invariance of $I_{\cal H}^{\beta_0}(\xi ,\eta)$ and  $I_{\Gamma}^{\beta_0} (\xi ,\eta)$}. \par
It can be readily verified, using (\ref{gxieta}), (\ref{full}) and the definitions of 
$I_{\cal H}^{\beta_0} (\xi ,\eta)$ and $I_{\Gamma}^{\beta_0} (\xi ,\eta)$.
\item
{\em $G_k$- covariance}\par
If $\ab'= \ub_k^{-1} (u,b,c) \ab \ub_k (u,b,c)$ is the operator transformed from $\ab$ by 
the   group representation (\ref{repgk}), it follows from (\ref{full}) that:
 \be
I_{\cal H}^{\beta_0} (\ab';\xi , \eta)= 
I_{\cal H}^{\beta_0} (\ab; e^u \,\xi + b , e^{ku} \, \eta +c )
\ee
On the other hand, if $\A'(t,f) = \A (e^ut+b +kce^{-(k-1)u} \, f^{k-1}, a^{-1} \, f)$, 
a direct computation leads to:
\be
I_{\Gamma}^{\beta_0} (\A';\xi , \eta )= 
I_{\Gamma}^{\beta_0} (\A; e^u \,\xi + b , e^{ku} \, \eta +c)
\ee

\item {\em 
Possibility of reconstruction of the operator $\ab$ and of the function $\A(t,f)$}.\par
In spite of the fact that the functions $\psi_{\beta_0}^{\xi \eta}(f)$ do not form a basis, 
it can be shown that the operator $\ab$ is completely characterized by its diagonal 
elements (\ref{diaggk}).  \par
The reconstruction of the function $\A (t,f)$ could be performed by inverting the Radon 
transform (\ref{radongk}).
\end{description}
These properties allow to base the correspondence rule on the
  identification:    
\be
I_{\cal H}^{\beta_0} (\xi , \eta)=I_{\Gamma}^{\beta_0} (\xi ,\eta) 
\ee
which can be seen as an extension of (\ref{iii}).
The resulting correspondence rule has the form:
\begin{eqnarray}
 \lefteqn{\A(t,f)   =}   \label{opverak} \\[4mm]
& f^{2r+2} {\displaystyle \int}_{\R} e^{2i\pi u \beta_0} \, 
e^{2i\pi (tf-\beta_0)(\lambda_k(u) - \lambda_k(-u))}\, 
A(f\lambda_k(u), f\lambda_k(-u)) 
    \,  (\lambda_k(u) \lambda_k(-u))^{r+1}  \; du &
\nonumber
\end{eqnarray}
\noindent  where the functions $\lambda_k(u)$ are defined by
\be
\lambda_k(u) = e^{u/2} 
\left(k {\displaystyle \frac{\sinh(u/2)}{\sinh (ku/2)}} \right)^{1/(k-1)}
\ee
Conversely the kernel of the operator $\ab$ can be expressed in terms of the symbol by:
\begin{eqnarray}
A (f_1,f_2) & = & \int e^{-2i\pi (f_1-f_2)(t-(\beta_0 /f))}  
\delta (f_1-f_2-k^{-1}f^{1-k} (f_1^k-f_2^k))  \label{averopk} \\[2mm]
& \times &
\left( \frac{f_1}{f_2} \right)^{-2i\pi \beta_0 } (f_1f_2)^{-r}f^{-k}|f_2^{k-1} -f_1^{k-1}| 
\A(t,f) \; dtdf 
\nonumber
\end{eqnarray}
The operator $\ab$ corresponding to this kernel  can be 
written  in Weyl's form as:
\be
\ab = \int_{\R^2} \widehat{\A}(k,\beta_0;u,v) e^{-2i\pi (u {\bos \beta} + v{\bos f})} \; dudv
\ee
\noindent where
\begin{eqnarray} 
\widehat{\A}(k,\beta_0;u,v) & \! \! \!   = \! \! \! &
e^u  \left|  \frac{1-e^{(k-1)u}}{(1-k) u} \right| (\lambda_k(-u))^{k+1} 
e^{2i\pi \beta_0( u-\lambda_k(u) + \lambda_k(-u))} \label{ktrans} \\[2mm]
& \! \! \!   \times \! \! \! & \int_{\Gamma} 
e^{2i\pi tf ( \lambda_k (u) - \lambda_k (-u))} \, 
 \, e^{2i\pi f(v/u) \lambda_k (-u)(e^u-1)}  
  \A(t,f) \; fdtdf
\nonumber
\end{eqnarray}
It may be noticed that the limit for $k=0$ of function $\lambda_k(u)$ is identical to $ \lambda (u)$ 
defined in (\ref{lambda}). In that case, the relation  
(\ref{ktrans}) becomes a Fourier transform with respect to $\beta=tf$ and $f$ and 
the rule (\ref{therule})-(\ref{af}) is recovered. In the general case, the Fourier 
transform (\ref{af}) is replaced by the $k$-dependent transform (\ref{ktrans}) for which 
Parseval's formula does not hold. As a consequence, the unitarity property 
(\ref{unitary}) cannot be obtained.\par
However, it is possible to recover  an expression for the scalar product of 
operators in terms of 
symbols. This is accomplished by associating two symbols with any operator $\ab$: 
the symbol $\A(t,f)$ defined by (\ref{opverak}) and  a dual symbol $\A_d(t,f)$ defined by:
\begin{eqnarray}
 \lefteqn{\A_d(t,f)   =}   \label{opverakdual} \\[4mm]
&  & f^{2r+2} {\displaystyle \int}_{\R} e^{2i\pi u \beta_0} \, 
e^{2i\pi (tf-\beta_0)(\lambda_k(u) - \lambda_k(-u))}\, 
A(f\lambda_k(u), f\lambda_k(-u)) 
    \,  \mu_d(u)
 \; du 
\nonumber
\end{eqnarray}
where 
\be
\mu_d(u) =  
(\lambda_k(u) \lambda_k(-u))^{r+1}
\left| {\displaystyle \frac{d}{du}} (\lambda_k(u)-\lambda_k(-u)) \right|
\ee
It can be observed that it is only in the case $k=0$ that:
\be
 {\displaystyle \frac{d}{du}} (\lambda_k(u)-\lambda_k(-u)) =1
\ee
leading to 
\be
\A(t,f)= \A_d(t,f)
\ee
For an arbitrary value of $k$, the  unitarity formula  is 
   replaced by:
 \be
\mbox{Tr} \; (\ab \bb^{\dagger}) = 
\int_{\Gamma} \A(t,f) {\cal B}_d^*(t,f)\; dtdf
\label{unitgk}
\ee
\noindent where $\A(t,f)$ is the symbol of $\ab$ and ${\cal B}_d(t,f)$ is the dual 
symbol of $\bb$.\par
These different correspondence rules may be of interest for the study of special operators. 
The case $k=-1$ where
\be
\lambda_{-1} (u) =e^{u/2}
\ee
 stands out.
The symbolic calculus corresponding to $k=-1$ and $\beta_0=0$ has been introduced directly  
  by A.Unterberger \cite{unterberger} who applied it in a mathematical context.
 \par
A Wigner function $\Pc_k^{\beta_0} (t,f)$ can be associated with each $k$-calculus. 
Because of the form of relation (\ref{unitgk}),  the Wigner function of 
signal $S(f)$ is defined as the dual symbol of the projector on the signal. 
We thus associate a time-frequency distribution with 
each value of $(k,\beta_0)$. When $\beta_0=0$, these distributions coincide with 
the generalized  passive distributions introduced in \cite{jmp}. The case 
$k=1$, which has not been treated explicitly here, can 
be deduced by continuity.
 
\section{Conclusion}
A correspondence rule between functions on the time-frequency half-plane 
and operators on the Hilbert space of positive-frequency signals has been obtained. The rule, 
which is said geometric,  
is entirely based on the study of the affine group and more specially 
on the representations of its subgroups in the two domains. It is obtained, 
relatively to each subgroup, by identification of the decomposition in invariant 
subspaces with the corresponding tomographic decomposition in the time-frequency 
half-plane. The construction is very natural and the result can be considered as an affine version 
of the Weyl rule introduced in quantum mechanics.\par
In the correspondence, a Lie algebra of hermitian operators is transformed into a 
Lie algebra of symbols, also called star algebra. The basic operations of this algebra 
are the star product and the star bracket of symbols. Formulas for these operations have been 
given. A special attention has been paid to the case where the star bracket reduces to a 
Poisson bracket. \par
The obtained symbolic calculus has been integrated into a family of calculi 
covariant by three-parameter groups $G_k$, $k\in \Rbig$, containing the affine group. In this 
family, the geometric   rule
  corresponds to the special value $k=0$ and is the only one to  ensure the unitarity 
property.
\par
\vspace{5mm}
\noindent 
{\bf Acknowledgments}
\par
\vspace{5mm}
\noindent The authors are grateful to D.Sternheimer for his constructive remarks on the manuscript.
\vfill\eject

\end{document}